\documentclass[12pt]{iopart}

\usepackage{iopams}
\usepackage{setstack}

\usepackage[dvips]{graphicx}

\begin{document}

\title[Resonance production in heavy ion collisions]{Resonance production in heavy ion collisions}

\author{Christina Markert}

\address{\dag\ Physics Department, Kent State University, Kent, OH 44242,
USA}

\begin{abstract}
Recent results of resonance production from RHIC at $\sqrt{s_{\rm
NN}} = $ 200 GeV and SPS at $\sqrt{s_{\rm NN}} = $ 17 GeV are
presented and discussed in terms of the evolution and freeze-out
conditions of a hot and dense fireball medium. Yields and spectra
are compared with thermal model predictions at chemical
freeze-out. Deviations in the low transverse momentum region of
the resonance spectrum of the hadronic decay channel, suggest a
strongly interaction hadronic phase between chemical and kinetic
freeze-out. Microscopic models including resonance rescattering
and regeneration are able to describe the trend of the data. The
magnitude of the regeneration cross sections for different inverse
decay channels are discussed. Model calculations which include
elastic hadronic interactions between chemical freeze-out and
thermal freeze-out based on the K(892)/K and
$\Lambda$(1520)/$\Lambda$ ratios suggest a time between two
freeze-outs surfaces of $\Delta \tau>$~4~fm/c. The difference in
momentum distributions and yields for the $\phi$(1020) resonance
reconstructed from the leptonic and hadronic decay channels at SPS
energy are discussed taking into account the impact of a hadronic
phase and possible medium modifications.
\end{abstract}

\section{Introduction}

In heavy ion collisions an extended hot and dense fireball medium
is created. The properties (mass, width, momentum distribution,
yield) of the produced resonances depend on the fireball
conditions of temperature and pressure. During the fireball
expansion the short lived resonances and their hadronic decay
daughters may interact with the medium. Two freeze-out surfaces
can be defined, chemical and thermal, representing the conditions
when inelastic and elastic interactions cease respectively. In a
dynamical evolving system produced resonances decay and may get
regenerated. Hadronic decay daughters of resonances which decay
inside the medium may also scatter with other particles from the
medium. For SPS and RHIC energies these are mostly pions. This
results in a signal loss, because the reconstructed invariant mass
of the decay daughters no longer matches that of the parent.
Leptonic decay daughters on the other hand are unaffected by the
nuclear medium due to their small interaction cross section. The
rescattering and regeneration (pseudo-elastic) processes for
resonances and their decay particles depend on the individual
cross sections and are dominant after chemical but before the
kinetic freeze-out. These interactions can result in changes of
the reconstructed resonance yields, momentum spectra, widths and
mass positions. Rescattering will decrease the measured resonance
yields while regeneration will increase them.

Microscopic model calculations attempt to include every step in a
heavy ion interaction in terms of elastic and inelastic
interactions of hadrons and strings. They are therefore better
able to describe the rescattering and regeneration of the
resonances from fireball interactions. The prediction of a
specific model (UrQMD) is a signal loss for some of the resonances
due to more rescattering than regeneration in the low momentum
region p$_{\rm T}<1$~GeV for the hadronic decay channels
\cite{ble02,ble02b}. Comparisons between the yield and momentum
spectra of the hadronic and leptonic decay channels can indicate
the magnitude of the rescattering and regeneration contribution
between chemical and thermal freeze-out. In order to try to
understand the medium effect during the evolution and expansion of
the hot and dense fireball, we compare resonance yields and
spectra (width and mass) from elementary p+p and heavy ion
collisions and the results from the leptonic and hadronic decay
channels. An observed difference may give an indication of
in-medium modification of resonance properties.

\section{Resonance Reconstruction}

The signal loss due to rescattering is caused by the method of
measurement, the invariant mass is not properly reconstructed if
one of the decay daughters rescatters with another particle of the
surrounding medium. All the resonances are reconstructed by the
invariant mass of the decay daughters. The decay candidates are
identified by different techniques, their energy loss (dE/dx),
energy or displaced vertex (V0-reconstruction). The resonance
signal is obtained by the invariant mass reconstruction of each
daughter combination and subtraction of the combinatorial
background calculated by mixed event or like-signed techniques.
The resonance ratios, spectra and yields are measured at
mid-rapidity for RHIC at $\sqrt{s_{\rm NN}} = $ 200 GeV and over
4$\pi$ for SPS at $\sqrt{s_{\rm NN}} = $ 17 GeV. The central
trigger selection for Au+Au collisions at RHIC takes the 5\% or
10\% and for Pb+Pb collisions at SPS the 5\% of the most central
inelastic interactions. The setup for the p+p interaction is a
minimum bias trigger.

\section{Resonance Yields}

\begin{figure}[htb]
 \centering
 \includegraphics[width=0.8\textwidth]{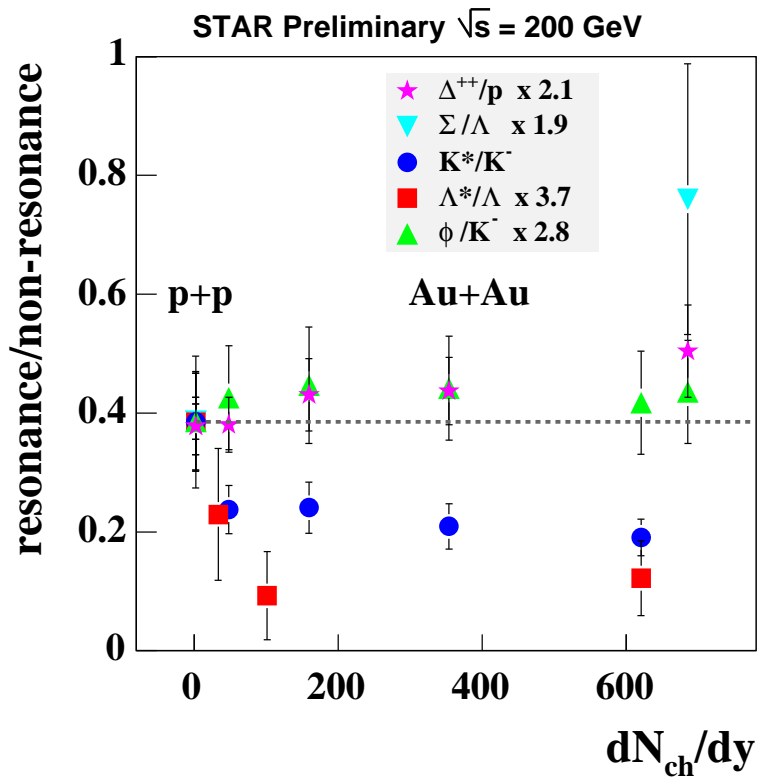}
\caption{Resonance/non-resonance ratios of $\phi$/K$^{-}$
\cite{ma04}, $\Delta^{++}$/p \cite{mar04qm}, $\rho/\pi$
\cite{fac04}, K(892)/K$^{-}$ \cite{zha04} and
$\Lambda$(1520)/$\Lambda$ \cite{gau04,mar03} for p+p and Au+Au
collisions at $\sqrt{s_{\rm NN}} = $ 200 GeV at mid-rapidity. The
ratios are normalized to the K(892)/K$^{-}$ ratio measurement in
p+p. Statistical and systematic errors are included.}
 \label{part}
\end{figure}

The resonance multiplicities at mid-rapidity for p+p and
peripheral to central Au+Au collisions at RHIC energies are
obtained for $\phi$(1020) \cite{ma04}, $\Delta(1232)^{++}$
\cite{mar04qm}, K(892) \cite{zha04} $\Lambda$(1520)
\cite{gau04,mar03} and $\Sigma$(1385) \cite{sal04}. In order to
compare different collision systems we normalize the yield to the
yield of the corresponding measured ground state particle. Under
the assumption that the Au+Au collision system is only a
superposition of p+p collisions we would expect the same
resonance/non-resonance ratio. Fig.\ref{part} shows the
resonance/non-resonance ratios normalized to the K(892)/K
measurement in p+p. The $\Lambda$(1520)/$\Lambda$ and the K(892)/K
ratio decreases from p+p to peripheral and central Au+Au
collisions.

The observed ratios can not be described by thermal model
predictions \cite{pbm01}, most likely because rescattering of the
decay daughters in the medium and regeneration are contributing to
the yield. If only rescattering occurs then the shorter lifetime
of the K(892) (4 fm/c) compared to the $\Lambda$(1520) (13 fm/c)
would result in a larger suppression for K(892)/K than for the
$\Lambda$(1520)/$\Lambda$ ratio. This implies that the
regeneration cross section is larger for the K+$\pi$ channel than
for the K+p channel. The $\phi$(1020)/K ratio is constant in all
collision systems within errors and can be described with the
thermal model, which is expected because only a small fraction of
the $\phi$(1020) are decaying inside the fireball due to the long
lifetime of the $\phi$(1020) (46 fm/c). The expected contribution
of rescattering for the short lived $\Delta$(1232) (1.7 fm/c) is
larger than that for the K(892) and the $\Lambda$(1520). However
the $\Delta$(1232)/p ratio does not decrease from p+p to Au+Au
collisions and is on the order of 41\% $\pm$ 22\% higher than the
thermal model prediction. This indicates a large cross section for
the regeneration of $\Delta$(1232) resonance in the p+$\pi$
channel. the $\Delta$(1232) can be re-created until T = 80-90 MeV
close to the kinetic freeze-out \cite{ble04}. The
$\Sigma$(1385)/$\Lambda$ ratio appears to follow the same trend as
the $\Delta$(1232)/p \cite{sal04}. This implies that the
$\Lambda$+$\pi$ regeneration cross section is nearly as high as
the p+$\pi$ regeneration cross section. From this observation we
can conclude that there is a ranking order of the cross section
for the different
regeneration processes: \\
$\sigma_{p+\pi}$ $\geq$ $\sigma_{\Lambda+\pi}$ $>$
$\sigma_{K+\pi}$ $>$ $\sigma_{K+p}$. The microscopic model
calculations (UrQMD) are able to reproduce the
resonance/non-resonance ratios in Au+Au collisions for most
resonances \cite{ble02,ble02b}. However the UrQMD prediction for
the $\Sigma$(1385)/$\Lambda$ ratio is in the order of 40\% $\pm$
20\% too high. In this calculation the assumption was made that
the $\Lambda$+$\pi$ regeneration cross section is the same than
for p+$\pi$. The trend of data would suggest that the
$\Lambda$+$\pi$ regeneration cross section is smaller than the
p+$\pi$ cross section.

\begin{figure}[htb]
 \vspace{0.5cm}
 \centering
 \includegraphics[width=0.5\textwidth,angle=-90]{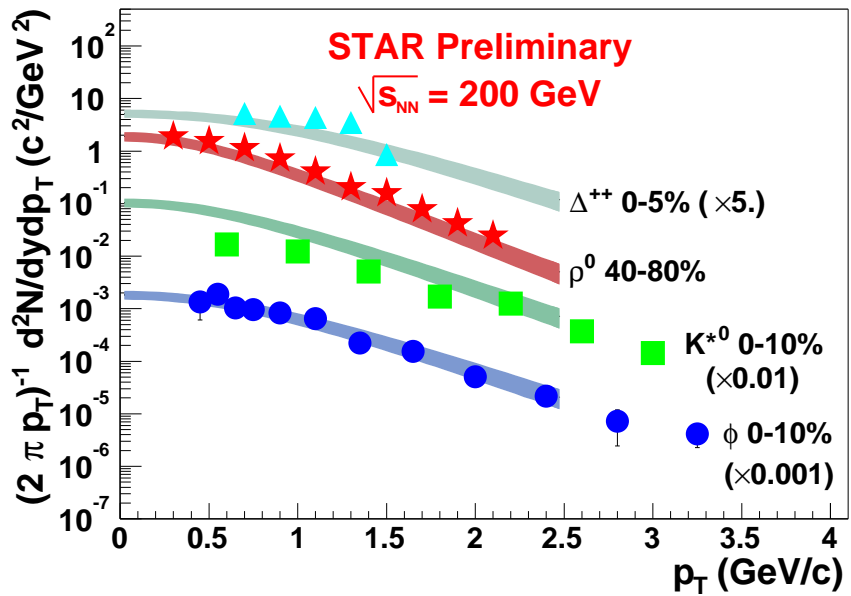}
\caption{Transverse momentum distribution of $\Delta^{++}$,
$\rho$, K(892) and $\phi$(1020) in central ($\rho$ peripheral)
Au+Au collisions from the STAR experiment at RHIC compared to
thermal model predictions \cite{flo04}.}
 \label{ptmodel}
\end{figure}

\section{Momentum Distribution}

Fig.~\ref{ptmodel} shows the momentum distribution of
$\Delta^{++}$, $\rho$, K(892) and $\phi$(1020) from central Au+Au
collisions ($\rho$ peripheral) from the STAR experiment at RHIC
compared to thermal model predictions from W. Florkowski. The
measured K(892) distribution deviates from the model predictions
in the low momentum region. This observation is consistent with
the UrQMD prediction of a signal loss due to rescattering in the
low momentum region. Based on the similarity in the trends between
the $\Lambda$(1520)/$\Lambda$ and the K(892)/K in Fig.~\ref{part},
one would also expect a signal loss in the low momentum region for
the $\Lambda$(1520) compared to the thermal model predictions. The
good agreement of the $\Delta^{++}$ momentum distribution with the
model indicats that the regeneration also takes place
predominantly in the low momentum region.

\begin{figure}[h]
\centering
 \vspace{0.5cm}
\includegraphics[width=0.55\textwidth]{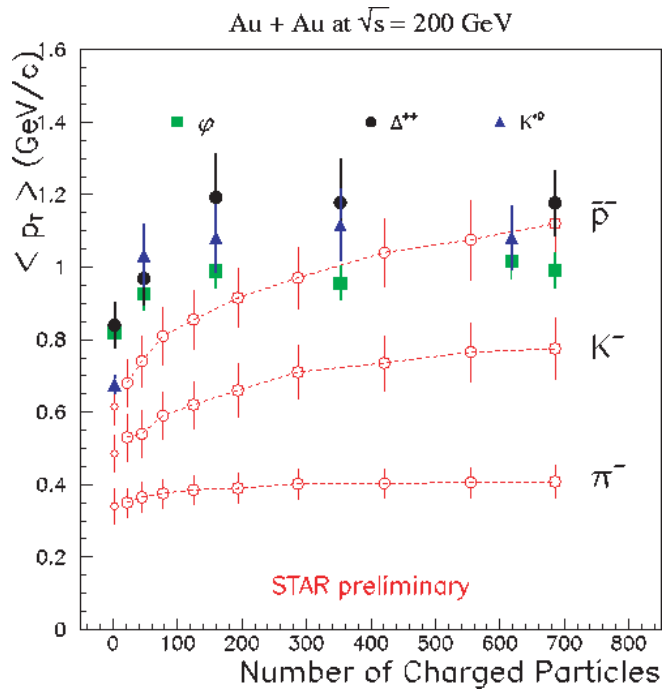}
 \caption{The $\langle$p$_{\rm T}$$\rangle$ for resonances
 and ground state particles in p+p and Au+Au collisions
 versus number of charged particles \cite{ma04,mar04qm,zha04,sal04}.}
 \label{resopt}
\end{figure}

This low momentum signal loss of resonances due to rescattering in
results in a higher inverse slope parameter and a higher
$\langle$p$_{\rm T}$$\rangle$. The STAR data from p+p and Au+Au
collisions at $\sqrt{s_{\rm NN}} = $ 200 GeV confirm this trend. A
strong increase $\langle$p$_{\rm T}$$\rangle$ for resonances is
observed from p+p to the most peripheral Au+Au measurement. The
same trend is not present for the ground state particles (see
Fig~\ref{resopt}) \cite{ma04,mar04qm,zha04}.

\section{Time Scale}

Depending on the length of the time interval between chemical and
kinetic freeze-out, $\Delta \tau$, the magnitude of the
suppression factor of the measured resonance will change due to
contributions from rescattering and regeneration. A model using
thermally produced particle yields at chemical freeze-out and an
additional rescattering phase, including the lifetime of the
resonances and decay product interactions within the expanding
fireball, can yield an estimated $\Delta \tau$
\cite{tor01,tor01a,mar02}. This model does not include
regeneration and therefore predicts a lower limit of the lifetime
between the two freeze-out surfaces. The two ratios K(892)/K and
$\Lambda$(1520)/$\Lambda$ are expected to have a larger
rescattering contribution. A $\Delta\tau$ $>$ 4~fm/c results if
chemical freeze-out occurs at 160 MeV.

\section{Leptonic and Hadronic Decay Channels}

In heavy ion collisions direct comparisons of the spectra and
yields obtained from leptonic and hadronic decay channels of a
single resonance may show the influence of the hadronic
interaction phase after chemical freeze-out. The $\phi$(1020) is
one of the resonances where we have measurements of the leptonic
and hadronic decay channel. At SPS energies the reconstruction of
the $\phi$(1020) in the different decay channels seemingly leads
to differing $\phi$(1020) kinematics and yields ($\phi$ puzzle).

\begin{figure}[htb]
\vspace{0.8cm}
 \centering
\includegraphics[width=0.6\textwidth]{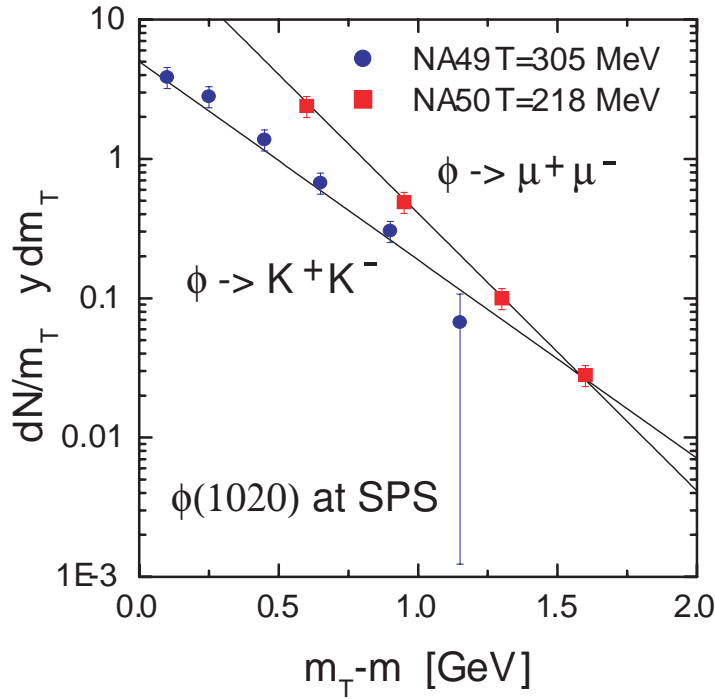}
\caption{Transverse momentum distribution of the hadronic decay
$\phi$(1020) $\rightarrow$ K$^{+}$ + K$^{-}$ from NA49
\cite{fri97} and the leptonic decay $\phi$(1020) $\rightarrow$
$\mu^{+}$ + $\mu^{-}$ from NA50 \cite{wil99}.}
 \label{phi}
\end{figure}

Fig.~\ref{phi} shows the transverse momentum distribution from the
hadronic decay $\phi$~$\rightarrow$~K$^{+}$~+~K$^{-}$ (NA49) and
the leptonic decay $\phi$~$\rightarrow$~$\mu^{+}$~+$\mu^{-}$
(NA50) \cite{fri97,wil99}. The inverse slope parameter from fits
to the momentum spectra, indicated as lines, are
T~=~305~$\pm$~15~MeV for hadronic decay and T~=~218~$\pm$~10~MeV
for leptonic decay. The extracted yield from the extrapolation of
the momentum spectrum of the leptonic decay is a factor of
4~$\pm$~2 higher than the one for the hadronic decay. Measurements
of the $\phi$(1020) reconstructed via the hadronic and leptonic
decay from CERES presented by A. Marin \cite{mari04} at this
conference confirm the NA49 results ($\phi$ $\rightarrow$ K$^{+}$
+ K$^{-}$) in terms yield and momentum distribution and the NA50
yield for the $\phi$ $\rightarrow$ $e^{+}$ + $e^{-}$ decay. First
results from NA60 experiment show an improved invariant mass
signal (significance $>$ 20) for the  $\phi$ $\rightarrow$
$\mu^{+}$ + $\mu^{-}$ channel \cite{dam04}, which should result in
a conclusive contribution to the $\phi$ puzzle at SPS.

Microscopic calculations (UrQMD) estimate a suppression of 20-30\%
of the $\phi$(1020) yield in the hadronic decay channel due to
rescattering of the kaon decay daughters in the low momentum
region p$_{\rm T}$~$<$~1~GeV \cite{ble02,ble02b}. The rescattering
is negligible for the leptonic decay due to the very low cross
section of interaction with the hadronic phase. Therefore the
lower signal in the low momentum region of the hadronic decay
(NA49) compared to the leptonic decay (NA50) is in agreement with
the model. However the signal loss of 20-30\% from the model
calculation is not sufficient to explain the factor of 4~$\pm$~2
in the measured yield of the data.

This allows for possible medium effects on the resonance
production which are likely to occur at an earlier stage, before
chemical freeze-out. Alternative calculations to describe in
medium modification of the $\phi$(1020) resonance were published
recently by K. Haglin and E. Kolomeitsev
\cite{hag04,hag04a,kol99}. Here the lifetime of the $\phi$(1020)
resonance is modified towards smaller lifetimes due to
modification of the spectral functions in the hot and dense
fireball and therefore more of the $\phi$(1020) resonances decay
inside the medium. This will introduce a larger signal loss due to
rescattering of the hadronic decay daughters.

\section{Feeddown from Resonances}

Finally I would like to conclude with a small remark. If we
interpret particle spectra of ground state particles we have to
take into account that a large fraction of the particles are
coming from resonance feeddown, as already pointed out by E.
Schnedermann et al. \cite{sch93}. For the proton we have 42\% from
$\Lambda$'s, 21\% from $\Delta$'s, and 11\% from $\Sigma^{0}$'s
(statistical model \cite{raf}). Therefore only 26\% of the protons
are primary produced protons. 35\% of the $\Lambda$'s are from
$\Sigma$(1385) and 20\% from $\Sigma^{0}$'s (statistical model)
decays. If we take the contribution of multiple rescattering and
regeneration processes during the expansion of the fireball source
into account, the number of primary particles will be further
reduced, because the regeneration does not necessarily involve the
actual resonance decay particles. Since the lifetimes of the
$\rho$ and $\Delta$(1232) are very short compared to the lifetime
of the fireball, we would expect a larger number of $\pi$'s and
protons coming from a $\Delta$(1232) decay than from higher mass
baryons. Therefore many $\pi$'s and protons are coming from a
later stage of the evolution of the fireball source and their
momentum distribution might be different from the primary produced
particles. Conclusions based on the momentum distributions of
particle spectra in terms of flow and freeze-out temperatures have
to take the contribution from resonance decays into account.



\end{document}